\documentstyle[12pt]{article}
\begin{document}
\setlength{\baselineskip}{0.27in}

\newcommand{\beq}{\begin{equation}}
\newcommand{\eeq}{\end{equation}}
\newcommand{\beqa}{\begin{eqnarray}}
\newcommand{\eeqa}{\end{eqnarray}}
\newcommand{\lsim}{\begin{array}{c}\,\sim\vspace{-21pt}\\<
\end{array}}
\newcommand{\gsim}{\begin{array}{c}\sim\vspace{-21pt}\\>
\end{array}}

\begin{titlepage}
{\hbox to\hsize{April 1995\hfill } }
{\hbox to\hsize{SPhT Saclay T95/043\hfill }}
{\hbox to\hsize{UM-TH-95-12\hfill hep-ph/xsses}}

\begin{center}
\vglue .6in
{\Large \bf
On the Universality of the Leading, $1/Q$ Power Corrections in QCD}
\\[0.5in]
\begin{tabular}{c}
{\bf R. Akhoury\footnotemark[1]}\\[.05in]
{\it CEA, Service de Physique Theorique, CE-Saclay}\\
{\it F-91191 Gif-Sur-Yvette, Cedex, France}\\[.05in]
{\bf V. I. Zakharov}\\[0.05in]
{\it Randall Laboratory of Physics}\\
{\it University of Michigan, Ann Arbor, MI 48109}\\[.15in]
\end{tabular}
\footnotetext[1]{On Sabbatical leave from the University of Michigan, Ann
Arbor}
\vskip 0.25cm
{\bf Abstract}\\[-0.05in]

\begin{quote}
We discuss $1/Q$ corrections
 to hard processes in QCD where $Q$ is a large mass parameter
like the total energy in $e^+e^-$ annihilation. The main problem we address
ourselves to is
whether these corrections to different processes (concentrating for
definiteness on the Thrust
and the Drell-Yan cross section) can be related to each other in a reliable way
so that the
phenomenology of the $1/Q$ corrections can be developed. We derive first the
relation valid
to lowest order using both the renormalon and finite-gauge-boson mass
techniques to check
its independence on the infrared cut- off procedure. We then argue
that the $1/Q$ corrections are due to soft gluons which factorize into a
universal factor
such that the lowest order relations are preserved in higher orders.
\end{quote}
\end{center}
\end{titlepage}
\newpage

{\bf 1.} Power like corrections, $(\Lambda_{QCD}/Q)^{\delta}$, where
$Q$ is a large mass scale provide important insight into non-perturbative
dynamics in QCD.
The best known example are the QCD sum rules \cite{svz} where the leading power
corrections
are proportional to the gluonic condensate, $\langle G^2 \rangle /Q^4$.
The knowledge of the power corrections in that case is based on the OPE.
Since OPE is essentially a Euclidean construction the limitation of this
approach is
that it is applicable only for physical quantities that can be obtained by
analytical continuation from
Euclidean space, such as the total cross section for
$e^+e^-\rightarrow hadrons $. The power like corrections can also be studied
\cite{vz}
by means of infrared renormalons since the latter are particular graphs
sensitive to large distances.
Thus, all the general properties are observed and the nonperturbative
contribution
results in an overall factor which can be large numerically.

The advantage of the renormalon technique is that it can be applied to any
infrared safe
quantity and it is by using this technique and its variations that $1/Q$
corrections were
discovered recently \cite{cs,bigi,bb,mw,webber,ks}.
One of the modifications of the renormalon technique
is introduction of finite gluon mass ($\lambda$)
in the evaluation of the lowest order radiative corrections
\cite{bigi,bbz,webber}. In this case one looks for $\alpha \lambda /Q$
corrections and, in particular, in reference
\cite{webber} an attempt was made to relate
the $1/Q$ corrections to different event shape variables.

{\bf 2.} Since $1/Q$ corrections are the leading ones in most cases, their
theory is of immediate
interest. Moreover, the very fact of existence of such corrections follows from
simple dimensional
estimates of the infrared contribution to one or another inclusive
quantity by means of perturbative graphs.
The non-perturbative effects bring then an unknown overall factor.
To develop a phenomenology of
$1/Q$ corrections one needs however to derive relations between these
corrections to different
processes in a reliable way.

In the absence of the OPE there are apparent difficulties in deriving such
relations.
In particular in the method
of introducing a non-zero $\lambda$ it is not at all clear how to go beyond the
leading order
because of the non-Abelian nature of the interaction of the gluons. Similarly,
in the renormalon
technique,(here we follow the approach of \cite{ks})
higher order corrections are unsuppressed because the coupling constant appears
to
be normalized at a low scale (see later discussion). Also, it is far from being
obvious
that relations between the $1/Q$ corrections obtained in one scheme of
introducing an infrared
cutoff will be valid universally.

These are the questions to which we address ourselves in this paper. For sake
of definiteness
we concentrate first on a relation between $1/Q$ corrections to Thrust and to
the Drell-Yan
process. It is straightforward to derive such a relation in lowest order in
$\alpha$ and
we check its validity  by deriving it in the two schemes mentioned above.
Next we discuss the origin of the $1/Q$ corrections and
argue for their universality. The key observation is that the same kinematical
region of soft
gluons is responsible for $1/Q$ correctins in Thrust and the Drell-Yan cross
section.
The soft part factorizes and the region of the large coupling does not affect
the relation between
the quantities. It is natural, then, to assume  then the non-perturbative
corrections due to the
large distances factorize as well. We conclude with discussing the other event
shape variables.

{\bf 3.} We proceed now to the computation of the $1/Q$ correction, to the
lowest nontrivial
order, to the Thrust variable. This quantity is defined as:
\beq
T~=~max_{\bf n}\frac {\displaystyle \Sigma_i|{\bf p}_{i,n}|}
{\displaystyle \Sigma_i |{\bf p}_i|}
\label{thrust}\eeq
where, ${\bf p}_{i,n}$ is the component of i-th particle momentum,
${\bf p}_i$ in the reference direction ${\bf n}$ which is chosen to maximize
the right hand side of
above. Thus, for a twojet event, $T=1$ and for a completely symmetric 3-jet
event $T=2/3$.

For $e^+e^-\rightarrow q\bar{q}$, $T=1$ and in the first order, i.e. for
$q\bar{q}g$
final state
\beq
T~=~max (x_1,x_2,x_3)
\eeq
with $x_i=2E_i/Q$ in the CM frame where the momenta of $q,\bar{q}$ and $g$
respectively are
denoted by $p_1,p_2,p_3$ and $Q$ is the total CM energy. In the invariant basis
\beq
Q^2x_i~=~Q^2-(Q-p_i)^2.
\eeq
For massless quarks and gluons one finds the range of the variables as:
\beq
0\le x_2\le1;~~1-x_2\le~x_1\le~1~and~x_1+x_2+x_3~=~2.
\eeq
For a nonzero gluon mass $\lambda$, the above range is modified to
\beq
0~\le~x_2~\le~1-{\lambda \over Q},~1-x_2-{\lambda \over Q}~\le~x_1~\le~
{1-x_2-\lambda/Q\over{1-x_2}}.\label{range}
\eeq
The cross section for real emission of $q\bar{q}g$ can
be calculated in a straightforward manner
in terms of $\sigma_0$, the Born cross section for $e^+e^-\rightarrow
q\bar{q}$.
For $\lambda =0$, it is:
\beq
{1\over {\sigma_0}}
{d\sigma\over {dx_1dx_2}}
{}~=~{\alpha_s\over {2\pi}}C_F{x_1^2+x_2^2\over {(1-x_1)(1-x_2)}}  .
\label{born}
\eeq
The calculation of the $1/Q$ correction to $\langle 1-T\rangle$ for nonzero
$\lambda$ has been reported in ref \cite{webber}, so we merely point out some
salient features.
{}From (\ref{born})
 one first computes the Thrust distribution $1/\sigma_0 d\sigma/dT$
by substituting the phase space region according to $x_1,x_2,x_3$
being the largest fraction and finding $1/\sigma_0\cdot d\sigma/dT$
in each region. Adding these,
gives the total thrust distribution from which $\langle 1-T\rangle$ is easily
obtained. The
result is \cite{webber}:
\beq
\langle 1-T \rangle_{1/Q}~=~-4C_F{\alpha_s\over {\pi}}
{\lambda \over Q}.
\eeq
We wish to emphasize that the contribution to the $1/Q$ term only comes from
max $(x_1,x_2)$
(i.e. the region of energetic quarks + antiquark only) and, in these regions,
only from terms linear in $\lambda$ coming from the phase space constraints
(\ref{range}).
It is therefore seen to be soft-gluon dominated, i.e. coming from
gluon energies of order $\lambda$.

At this point, to set the stage for subsequent developments,
it is appropriate to
give precise definitions of what we call soft and what we call the hard
collinear
regions throughout this paper. If we denote a typical momentum by $k$, and if
by
$\delta$  we denote a small parameter whose vanishing give infrared sensitivity
(For example, $\delta$ could be $1-z$, the longitudinal fraction of the gluon
energy when we consider the $\tau\rightarrow 1$ limit in inclusive
leptoproduction)then
the two regions are defined thus.
\begin{eqnarray}
Soft: k_+ \sim k_- \sim k_{\perp} = {\cal O}(Q\delta).~~~~~~~~~~~~~\nonumber\\
Hard~ collinear: k_+ ={\cal O}(Q\delta),~
k_{\perp}={\cal O}(Q\sqrt{\delta}),~ k_-={\cal O}(Q).
\end{eqnarray}
In the Hard collinear case we take the original hard particle to be moving in
the
$''-''$ direction.

In the second method of evaluating the $1/Q$ corrections to thrust, it will
be convenient to introduce the variable "spherocity", $S$ \cite{gm}
\beq
S~=~min\left({4\over {\pi}}\right)^2
\left( \frac {\displaystyle \Sigma_i{\bf p}_{\perp}^i}
{\displaystyle \Sigma_i|{\bf p}|_i}\right)^2
\eeq
where, ${\bf p}_{\perp}^i$ is the transverse momentum with respect to the
minimum direction.
Thus $S=0$ for 2-jet event. For a three parton final state
\beq
S~=~{64\over {\pi^2T^2}}(1-x_1)(1-x_2)(1-x_3).
\eeq
If by $x_{\perp}={2k_{\perp}\over Q}$ we denote the fractional transverse
momentum of either parton in
opposite hemisphere of the fastest parton, then
$x_{\perp}^2={\pi^2\over 16}S$. Hence
\beq
{k_{\perp}^2\over {Q^2}}~=~{1\over {T^2}}(1-x_1)(1-x_2)(1-x_3).
\eeq
Furthermore, one has the limits:
\beq
{(1-T)^2(2T-1)\over {T^2}}~\le ~{k_{\perp}^2\over {Q^2}}~\le~{1-T\over 4}.\eeq
We next change variable: $x_1,x_2 \rightarrow T, k_{\perp}^2/Q^2 = x$
and from (\ref{born}) we find for the double differential cross section
\beqa
{1\over {\sigma_0}}{d\sigma \over {dxdT}}~=~
{\alpha_s\over {2\pi}}C_F
{T\over {(1-T)(1-4x/(1-T))^{1/2}}}~~~~~~~~~ \nonumber\\
 \left( {2(T^2+\Delta_+^2)\over {(1-T)(1-\Delta_+)}}
+{2(T^2+\Delta_-^2)\over {(1-T)(1-\Delta_-)}}+
{(2-T-\Delta_+)^2+\Delta_+^2\over {(T+\Delta_+-1)(1-\Delta_+)}}
+{(2-T-\Delta_-)^2+\Delta_-^2
\over{(T+\Delta_-1)(1-\Delta_-)}}\right)
\eeqa
where,
\beq
\Delta_{\pm}~=~1-{1\over 2}T\left(1\pm\sqrt{1-{4x\over {1-T}}}\right).
\eeq
Integrating over $x$ gives the well known expression \cite{altarelli}
\beq
{1\over {\sigma_0}}{d\sigma\over {dT}}~=~
{\alpha_s\over {\pi}}C_F
\left({4\over {T(1-T)}}ln{2T-1\over{1-T}}+6ln{2T-1\over {1-T}}-
{3(3T-2)(2-T)\over {1-T}}\right).
\eeq
we are interested in
\beq
\langle 1-T \rangle~=~
\int_{2/3}^1dT\int_{{(1-T)^2(2T-1)\over {T^2}}}^{1-T\over 4} dx
{1\over \sigma_0}
{d\sigma\over{dxdT}}(1-T).
\eeq
To get terms linear in $Q$, we interchange the integration procedure
\beq
\int_{2/3}^1dT\int_{{(1-T)^2(2T-1)\over{T^2}}}
^{{1-T\over 4}}dT~\rightarrow
\int_{x_1}^{x_2}dx\int_{T_1}^{T_2}dT
\eeq
and the new limits $x_1,x_2$ and $T_1,T_2$ are
\beq
x_1=0,~x_2={1\over 12};~~T_1=1-\sqrt{x}+...,~~T_2=1-4x+...~~,
\eeq where $T_1$ and $T_2$ are given as an expansion in $x$. Proceeding thus,
we find
that the only contribution to the $1/Q$ term
comes from the lower limit of $T$ integration: i.e. for $T\rightarrow 1$ and
the soft gluon region.

A contribution to $<1-T>_{1/Q}$ ha the form:
\beq
\langle 1-T\rangle_{1/Q}
{}~=~{\alpha_s\over \pi}C_F\int_0^{1/12}{dx\over x}
\int_{1-\sqrt{x}}^{1-4x}dT{4\over T}{1\over (1-{4x\over 1-T})^{1/2}}.
\eeq
Precisely this term gives the ${ln(1-T)\over 1-T}$ behaviour in the thrust
distribution
above. Next, to find the position of the leading infrared renormalon we allow
$\alpha_s$ to run according to $\alpha \rightarrow \alpha_s(k_{\perp}^2)$
(see, e.g., ref \cite{bassetto}) and in the above integral expand in
$k_{\perp}/Q$. The leading term is
\beq
\langle 1-T\rangle_{1/Q}~=~{2\over Q}
\int_0^{{1\over 12}Q^2}
{dk^2_{\perp}\over k^2_{\perp}}k_{\perp}
\left(C_F{\alpha_s(k_{\perp}^2)\over \pi}\right)
\approx
{4\over Q}\int_0^{\sim Q} dk_{\perp}
\left({\alpha_s(k^2_{\perp})\over \pi}C_F\right).
\eeq

The coefficient of $4{\alpha_s\over \pi}C_F$ is the same as for the $\lambda\ne
0$ calculation,
if we identify $\lambda$ with:
\beq
\lambda \sim \int_0^{\sim Q} dk_{\perp}
\left({\alpha_s(k^2_{\perp})\over \pi}C_F\right).
\eeq

The leading IR renormalon at $1/2b_0$ can be explicitly seen by introducing the
representation
\cite{ks}
\beq
\alpha_s(k^2_{\perp})~=~\int_0^{\infty}
d\sigma exp(-\sigma (b_0 ln{k^2_{\perp}\over \Lambda^2}))
\eeq
above and doing the $k_{\perp}$ integration to get a pole at $2\sigma b_0 = 1$.

We next turn to the $1/Q$ correction to the  Drell-Yan process
which was discussed in \cite{cs,ks}. Following \cite{ks}, the quantity of
interest is
the Drell-Yan cross section ${d\sigma_{DY}(\tau,Q^2)/dQ^2}$ normalized
to the structure functions of deeply inelastic scattering $F_{DIS}(x,Q^2)$.
In particular, we will be interested in the large $N$ moments of this
which is known to factorize to all orders \cite{sc}
Thus:
\beq
\label{dyt}
lim_{N\rightarrow \infty}{
\int_0^1 d\tau\tau^{N-1}
{d\sigma_{DY}(\tau ,Q^2)\over
dQ^2}~=~\sigma_0R(\alpha_s(Q^2))E(N,Q^2)F_{DIS}^2(N,Q^2)}
\eeq
with $\tau =Q^2/s$, $Q^2$ is the invariant mass of the dilepton pair and
$R(\alpha_s(Q^2)) $is a finite function as $N\rightarrow \infty$.
In the above, $E(N,Q^2)$ is the contribution of soft gluons only.
For large $N$ we are in the kinematical domain $\tau\rightarrow 1$ and here,
$E(N,Q^2)$ has
large perturbative corrections which in the lowest order are of type $log^2N$
and $log N$
coming from the region of soft gluons. It is well known \cite{sc} that these
exponentiate.
In $E(N,Q^2)$, the dominant term at large $N$ in the lowest order is
(to detect the renormalon we need to have the coupling
running so that we have let $\alpha_s\rightarrow \alpha_s(k^2_{\perp})$, which
is the appropriate scale for
$\alpha_s$, \cite{bassetto,sc}):
\beq
\label{threestar}
1-2\int_0^1dz{z^{N-1}-1\over 1-z}
\int_{(1-z)^2Q^2}^{(1-z)Q^2}{dk^2_{\perp}\over k^2_{\perp}}
{\alpha_s(k^2_{\perp})\over \pi}C_F.
\eeq
By interchanging the $k_{\perp}$ and $z$ integrations we can identify the $1/Q$
behaviour which was found in \cite{ks}
\begin{eqnarray}
\label{dyq}
E(N,Q^2) \sim
{2(N-1)\over Q}\int_0^{Q^2}
{dk^2_{\perp}\over k^2_{\perp}}\int_{1-k_{\perp}/Q}^{1-k^2_{\perp}/Q^2}dz
\left(C_F{\alpha_s(k^2_{\perp})\over \pi}\right)  \sim\nonumber\\
{2(N-1)\over Q}\int_0^{Q^2}
{dk^2_{\perp}\over k^2_{\perp}}k_{\perp}\left(C_F{\alpha_s(k^2_{\perp})\over
\pi}\right).
\end{eqnarray}
The coefficient of the term proportional to $(N-1)$
comes from real emission diagram and is exactly the same as the $1/Q$
corrections
in $\langle 1-T \rangle $ and is again related to the soft behaviour as
$\tau\rightarrow 1$.
This in turn can be traced back to the maximum allowed transverse momentum in
deeply inelastic
scattering and in Drell-Yan:
\beq
(k^2_{\perp})^{DIS}_{max}~=~{s\over 4}~=~{Q^2(1-z)\over 4z}
\eeq
and \beq
(k^2_{\perp})^{DY}_{max}~=~{(s-Q^2)^2\over 4s}~=~{Q^2(1-z)^2\over
4z}~=~(1-\tau)^2{s\over 4}.
\eeq
The $1/Q$ corrections come from the lower limit of the $z$ integration in
the first of equation
(\ref {dyq})
which is governed by the Drell-Yan soft dynamics as discussed above.
Further, the contributing region is the one where the invariant mass of the
gluon radiation
is vanishingly small, of order $(1-\tau )s$ as $\tau\rightarrow 1$.

The same $1/Q$ term can also be obtained by the method of introducing a gluon
mass
$\lambda $. In this case again the only contribution to the $1/Q$ term
comes from the
specification of the integration boundaries.
The coefficient of this term,linear in gluon mass, in the corresponding
expression for
$E(N,Q^2)$
is identical to the result using the other procedure
with the same identification
of $\lambda$ as for the case of Thrust.Furthur,
it is easily checked
to be coming from the region of soft gluons alone.

{\bf 4.} A few comments are in order concerning these results. Although we have
checked the
independence of the relation between the $1/Q$ terms on the choice of infrared
parameter to lowest
order, we should look at the problem in higher orders as well.
The point is that higher orders are not suppressed since $\alpha_s$ occurs
normalized at a low scale.
To demonstrate the lack
of suppression of higher orders within the renormalon technique, we notice
that in the leading log approximation
\beq
\alpha^2_s(k^2_{\perp})~=~\left( \Lambda_{QCD}{d\over 2b_0d\Lambda_{QCD}}
\right) \alpha_s(k^2_{\perp}).
\eeq
Since the renormalon contribution is linear in $\Lambda_{QCD}$
we conclude that $\alpha_s^2(k^2_{\perp})$.
contribution is not suppressed. Thus we are led to investigate the
origin of the universality of the $1/Q$ corrections obtained in the lowest
order.

Let us discuss the nature of the higher order corrections to the IR safe, event
shape variables $y$
that vanish in the limit of the 2-jet process (like  $1-T$ in particular).
{}From the lowest order analysis we have seen that the $1/Q$ corrections
come from the region $T\rightarrow 1$ and of energetic quarks (antiquarks) and
soft
gluons. Indeed, we have seen that as far as these $1/Q$ corrections are
concerned
we have in the CM frame
\beq
1-T~\approx~{M_q^2\over Q^2}+{M_{\bar{q}}^2\over Q^2}
\eeq
where $M_q^2\approx 2p_q\cdot k_i$ in the limit of soft gluons.
This generalizes straightforwardly to all orders: Let us first argue
for the dominance of soft gluons in $1-T$ in higher orders. In the next order,
the infrared
sensitive contribution will come if the additional gluon is either soft or
hardcollinear,
real or virtual. Since $(1-T)$ is an IR safe quantity, the hard collinear
gluon will cancell
against the virtual process and the only IR sensitive contributions that
will survive are
those involving  soft gluons. Continuing in this manner, it is seen that to all
orders
the infrared sensitive contributions that will survive are those involving the
soft gluons alone.
In fact, to all orders, the IR sensitive contributions to $1-T$ can come only
from the
region $T\rightarrow 1$ , where we can write
\beq
\label{alpha}
1-T~\approx~{M_1^2\over Q^2}+{M_2^2\over Q^2}
\eeq
Here we have considered two hemispheres, $H_1$ and $H_2$,
divided by a plane perpendicular
to the thrust axis, defined $P_1$ and $P_2$ to be the vector sum of the momenta
in these
and denoted by $M_1$ and $M_2$ the corresponding invariant masses. It is then
easy to
see that for small $1-T$ the above is true. In principle, both soft and hard
collinear
gluons contribute towards small invariant masses, however the latter are
excluded
since they would cause a collinear divergence in $1-T$. Indeed, it is well
known that
the collinear divergences cancell in the total cross-section when we sum over
final
degenerate states. The same is true for $<T>$ since the states degenerate with
the
hard collinear gluon have the same value of $T$. With  this in mind we may
write in the
limit of soft gluons with momenta $k$,
\beq
M_i^2= \sum_{k_j^2\in H_i} m_j^2
\label{invmass}
\eeq
with, $m_j^2=2k_jP_i$.
The next step in the argument is that the soft gluons factorize
in the two jet process under consideration and hence to all orders we may write
\beq
\label{tee}
{<1-T>}_{1/Q} = R_T(\alpha_s(Q^2)) E_{soft}(Q)
\eeq
$E_{soft}(Q)$ comes from soft gluons alone and can be expressed in the form
\beq
\label{esoft}
E_{soft}(Q) = 1/Q \int_{0}^{Q^2}{ dk_{\perp}^2\over k_{\perp}^2} k_{\perp}
\gamma_{eik}(\alpha_s(k_{\perp}^2))
\eeq
$\gamma_{eik}(\alpha_s(k_{\perp}^2))$ contains contributions from all orders
and can be obtained
in perturbation theory from the eikonal approximation. It has the perturbative
expansion:
\beq
\gamma_{eik}(\alpha_s)= {\alpha\over \pi}C_F+({\alpha\over \pi})^2{1\over
2}C_FK+\ldots .
\eeq
with,
\beq
K = C_A(67/18-\pi^2/6)-10/9{T_R}{N_f}.
\eeq
$R_T$ is not IR sensitive and can be
obtained as an expansion in $\alpha_s(Q^2)$.

The above arguments emphasize that the origin of the $1/Q$ terms lies in the
soft gluon radiation.
The same is true for other shape variables like the $C$ parameter, which
are infrared safe and which vanish in the two jet limit.
In fact all of these quantities,
generically labelled as $y$ earlier are
governed by the same soft dynamics as they vanish \cite{ert}.
Thus we have in general:
\beq
\label{y}
{<y>}_{1/Q} = R_y(\alpha_s(Q^2)) E_{soft}(Q)
\eeq
In the above the quantities $R_y$ are $y$ dependant, but the soft functions
$E_{soft}$ are universal.

We next turn to a discussion of Drell-Yan (inclusive lepton pair production)
for large $\tau$,
 in higher orders of perturbation theory. As
discussed earlier, the factorization in the $\tau \rightarrow 1$
limit has been
established as given by Eq.(\ref{dyt}) , with $E(N,Q^2)$ , purely a soft
function. The $1/Q$
corrections to all orders is of the form \cite{ks}:
\beq
\label{dyqq}
{1 \over F_{DIS}^2 (N,Q^2) \sigma_0}lim_{N\rightarrow \infty}{
\int_0^1 d\tau\tau^{N-1}
{d\sigma_{DY}(\tau ,Q^2)\over dQ^2}~=~(N-1)R(\alpha_s(Q^2))E_{soft}(Q)}
\eeq
The equality on the RHS is, of course, only for the $1/Q$ terms.
In the limit $\tau \rightarrow 1$ , the invariant mass of
the quark gluon system which also has the form of Eq.(\ref{invmass}) , becomes
vanishingly small
. For the quantity $<1-T>$, we have a similar situation expressed by
Eq.(\ref{alpha}), in the
limit of vanishing invariant mass. In fact, the large corrections due to the
soft gluons will be the same, in the sense that the functions $E_{soft}$ are
universal hence the same in Eq.(\ref{esoft}) and Eq.(\ref{dyqq}). In \cite{ks}
$\gamma_{eik}$
is called the cusp anomalous dimension.
The factor of $N-1$ in Eq.(\ref{dyqq}) is what is nescessary to
make the correspondance between the two equations complete at $\tau \rightarrow
1$
since the invariant mass of the quark gluon system ,
is proportional to $1-\tau$ for the Drell-Yan case. Because of the relation
between
$<1-T>$ and the other event shape variables, we see that in general,
in, Eq.(\ref{y}) and Eq.(\ref{dyqq}),
the function $E_{soft}$ will be universal, whereas the perturbatively
calculable (for large $Q^2$)
infrared insensitive quantities, $R_y(\alpha_s(Q^2))$, and $R(\alpha_s(Q^2))$
 may be different.

 Apart from the situation discussed above, the same $\gamma_{eik}$ also
determines the
 Infrared behaviour of the quark form factor and the so called velocity
dependant
 anomalous dimensions in Heavy Quark Effective Theory \cite{rk}. The
universality of
 the anomalous dimensions of the effective currents in HQET was first
noted in \cite{fg}.

Up until now we have only discussed the situation in perturbation theory. We
now take
perturbative calculations with $\alpha_s$ normalized at low momenta ( as in the
functions $E_{soft}$) as representative of large distance dynamics in QCD,
including
non perturbative effects. Thus it is natural to expect a factorization similar
to that derived perturbatively
  for the full theory as well. If so then the relations discussed between the
$1/Q$ terms in
inclusive lepton pair production and the event shape varibles $<y>$,
will be valid here also.

In conclusion, we have presented arguments that a single non perturbative
parameter describes
the $1/Q$ corrections in Drell Yan and certain event shape variables. The same
nonperturbative
parameter makes its appearance also in heavy quark physics.
All of these quantities
are related to this universal parameter through overall normalizations.
Thus theory of the
$1/Q$ corrections seems to be ripe for the confrontation with experiments.

We would like to thank the Service de Physique Theorique, CE-Saclay for
hospitality.This work
was supported in part by the US Department of Energy.

\end{document}